\documentclass[aip,jcp,superscriptaddress,reprint]{revtex4-1}
\usepackage{epsfig}
\usepackage{amsmath}
\usepackage{amsfonts}
\usepackage{graphicx}
\usepackage{fancybox}
\usepackage{subfigure}
\usepackage[usenames,dvipsnames]{color}
\usepackage[titletoc]{appendix}
\usepackage{sidecap}

\newcommand{\degree}{^{\circ} }

\newcommand{\rb}{\mathbf{r}}
\newcommand{\sbb}{\mathbf{s}}

\newcommand{\xb}{\mathbf{x}}

\newcommand{\kb}{\mathbf{k}}
\newcommand{\avg}[1]{\left<#1\right>}
\newcommand{\len}[1]{\left|#1\right|}
\newcommand{\brac}[1]{\left[#1\right]}

\newcommand{\para}[1]{\left(#1\right)}

\newcommand{\gradr}{\nabla_{\rb}}
\newcommand{\gradk}{\nabla_{\kb}}

\newcommand{\laplac}{\vec{\nabla}^2}

\newcommand{\rhoqs}{\ensuremath{\rho^{q\sigma}}}
\newcommand{\rhoq}{\ensuremath{\rho^q}}

\newcommand{\rhoG}{\ensuremath{\rho_G}}

\newcommand{\frho}{\hat{\rho}}
\newcommand{\frhoqs}{\hat{\rho}^{q\sigma}}
\newcommand{\frhoG}{\ensuremath{\hat{\rho}_G}}

\newcommand{\V}{\ensuremath{\mathcal{V}}}

\newcommand{\Mb}{\ensuremath{\mathcal{M}}}
\newcommand{\Mbs}{\ensuremath{\mathcal{M}}^\sigma}

\newcommand{\Rbar}{\overline{\textbf{R}}}

\newcommand{\hermite}[2]{\mathbf{H}_{#1}\para{#2}}

\newcommand*{\citen}{}
\DeclareRobustCommand*{\citen}[1]{%
  \begingroup
    \romannumeral-`\x 
    \setcitestyle{numbers}%
    \cite{#1}%
  \endgroup
}

\begin{document}


\title{Hydrophobicity Scaling of Aqueous Interfaces by an Electrostatic Mapping}


\author{Richard C. Remsing}
 \affiliation{  Institute for Physical Science and Technology, 
  Department of Chemistry and Biochemistry, and
Chemical Physics Program, University of
  Maryland, College Park, MD 20742}
\affiliation{Department of Chemical and Biomolecular Engineering, University of Pennsylvania, Philadelphia, PA 19104}
\email[]{rremsing@seas.upenn.edu}
 
  \author{John D. Weeks}
    \email[]{jdw@umd.edu}

  \affiliation{  Institute for Physical Science and Technology, 
  Department of Chemistry and Biochemistry, and
Chemical Physics Program, University of
  Maryland, College Park, MD 20742}



\begin{abstract}
An understanding of the hydrophobicity of complex heterogenous molecular assemblies is crucial to characterize and predict interactions
between biomolecules.
As such, uncovering the subtleties of assembly processes hinges on an accurate classification of the relevant interfaces involved,
and much effort has been spent on developing so-called ``hydrophobicity maps.''
In this work, we introduce a novel electrostatics-based mapping of aqueous interfaces that focuses on
the collective, long-wavelength electrostatic response of water to the presence of nearby surfaces.
In addition to distinguishing between hydrophobic and hydrophilic regions of heterogenous surfaces,
this electrostatic mapping can also differentiate between hydrophilic regions that polarize nearby waters in opposing directions.
We therefore expect this approach to find use in predicting the location of possible water-mediated hydrophilic interactions,
in addition to the more commonly emphasized hydrophobic interactions that can also be of significant importance.
\end{abstract}

\keywords{ Protein Hydration, Hydropathy Scales, Self Assembly, Polarization, Dielectric Response, Hydrogen Bonding }

\maketitle

\raggedbottom

\section{Introduction}

The formation of complex mesoscopic structures is generally driven by the self assembly of nanoscale building blocks.
Oftentimes, the assembly of individual units into a larger complex is driven by strong, direct interactions between the chemical constituents~\cite{Solomon:2011,Sheinerman:2000aa,Bishop:2009aa,Whitesides:2002aa}.
 In many cases, however, the assembly of molecular building blocks is triggered by the solvent and not by direct interactions between the individual units.
 For example, water-mediated interactions between nonpolar entities, known as hydrophobic effects~\cite{DewettingRev,ChandlerNatureReview},
induce the folding of proteins by minimizing the contact area between hydrophobic amino acid residues and water at the surface of the protein~\cite{Dill:1990,Dobson:2003,Mirny:2001,Thirumalai:2010aa}.
Both hydrophobic and electrostatic interactions are also thought to play a role in protein-protein binding processes~\cite{Jones:1996,DeLano:2002aa,Sheinerman:2000aa}.
Driven by such applications, many researchers have proposed empirical hydrophobicity scales
for characterizing and predicting the relative hydrophobicity of complex surfaces. 
In general, these approaches fall into one of two classes, \emph{surface-based} and \emph{water-based} hydrophobicity scales. 

Surface-based hydrophobicity scales~\cite{Kyte:1982aa,Zhou:2003,Hua:JPCB:2007,Kapcha:2014} seek to predict the relative hydrophobicity
of a water-surface interface from the properties of the surface alone.
Such mappings have generally been used to predict the properties
of proteins and other complex biological surfaces.
Classic hydrophobicity scales like the Kyte-Doolittle scale~\cite{Kyte:1982aa}
assign a value of relative hydrophobicity to each macromolecular building block or
residue.

Another residue-based scale was introduced by Berne and coworkers
to characterize the binding regions of protein surfaces involved
in protein-protein interactions, and hundreds of surfaces were examined to predict if
a dewetting transition could occur upon protein complex formation~\cite{Hua:JPCB:2007}.
This analysis indeed successfully uncovered several binding processes in which a dewetting transition
was likely to occur, as confirmed by molecular dynamics simulations with explicit water molecules~\cite{Hua:JPCB:2007}.

More recent approaches have demonstrated that incorporation of atomic-level details may
provide an even better characterization of surfaces than classifying
an entire residue as hydrophobic or hydrophilic. 
In particular, Kapcha and Rossky (KR)
have exploited the assignment of partial charges in atomistic models
of proteins to classify individual atomic units as hydrophobic or hydrophilic
on a binary scale~\cite{Kapcha:2014}.
Predictions from these KR maps are in much better agreement with more detailed results
than those from residue-based methods and
have been used to characterize a wide range of protein surfaces and binding pockets~\cite{Kapcha:2014}.

Despite these predictive successes, surface-based hydrophobicity scales
only implicitly account for the influence of surface topography and the chemistry of nearby atomic units on the hydrogen bond network of water.
In light of these limitations, new methods have been developed for providing a direct characterization of
the nonlocal response of water to complex material surfaces.
In particular, recent work has focused on the nature of density fluctuations
in bulk water and at aqueous interfaces~\cite{Godawat:PNAS:2009,Patel:JPCB:2014,Patel:PNAS:2011,Jamadagni:ARCB:2011,Acharya:Faraday:2010,Patel:JSP:2011,Patel:JPCB:2012}.

Density fluctuations are intimately related to solvation thermodynamics
through potential distribution theory, which can be used to relate the solvation free energy
of a hard object to the probability of observing a cavity or volume of equal shape
and size being empty~\cite{Widom:1963,HummerInfoTheory,GardeEntropy,PDTBook}.
A large cavity or extended hydrophobic surface disrupts the hydrogen bond network and density fluctuations are enhanced relative to those in the bulk or near a hydrophilic surface.
This disruption of the network is the key physical feature characterizing large scale hydrophobic effects, and this makes it easier to create a cavity near
a hydrophobic surface than a hydrophilic one.
Previous work has exploited this fact, and used the solvation free energy of hard shaped objects or cavities near a surface
as a measure of its hydrophobicity~\cite{Godawat:PNAS:2009,Patel:JPCB:2014,Patel:PNAS:2011,Jamadagni:ARCB:2011,Acharya:Faraday:2010,Patel:JSP:2011,Patel:JPCB:2012}.
 
Because this approach focuses on the response of water to a substrate,
density fluctuation-based mappings can account for both chemical and topographical complexities of a surface.
However performing such calculations, especially for large asymmetric volumes, can be
computationally intense, often requiring several simulations with advanced umbrella-sampling techniques to obtain the solvation free energy of a single volume.
Moreover, this type of mapping depends on the size, shape, and position
of the probe volume to be emptied and in the absence of a general theory, different choices can lead
to ambiguities in the estimate of hydrophobicity.
While fluctuation methods very successfully discriminate between hydrophobic and hydrophilic surfaces, little distinction is seen between
surfaces with similar hydrophilicity but different chemistries.

In this work, we introduce a simple approach to characterize chemically and topographically complex materials based
on the collective, long-wavelength electrostatic response of water to such surfaces.
Excluded volume constraints and the partial surface charges exploited in the KR scale can produce distortions and disruptions of the hydrogen bond network.
Extended hydrophobic regions can break hydrogen bonds, while polar regions can locally distort and pin bonds in the network.
This complicated nonlocal response rearranges the molecular dipoles in water and is encoded in the electrostatic potential.  

As we will see, the long-wavelength component of the induced potential readily distinguishes between hydrophobic and hydrophilic surfaces, 
and also provides additional insight into the net polarization of water at a polar interface.
This is used to differentiate between hydrophilic surfaces with similar contact angles but different chemical structures.
We anticipate that this computationally simple method of analysis, which requires only a single equilibrium simulation, will find use
as a tool to efficiently characterize complex surfaces with nanoscale chemical and topographical patterning, like the surface of proteins involved in complex formation.
However, the forces generating the distortions of the hydrogen bond network are by no means purely electrostatic, and direct physical connections to macroscopic
measurements of hydrophobicity like the contact angle or the solvation free energy of cavities will require more general treatments.

\section{Electrostatic response at uniform planar surfaces}

\begin{figure}
\centering
\includegraphics[width=0.48\textwidth]{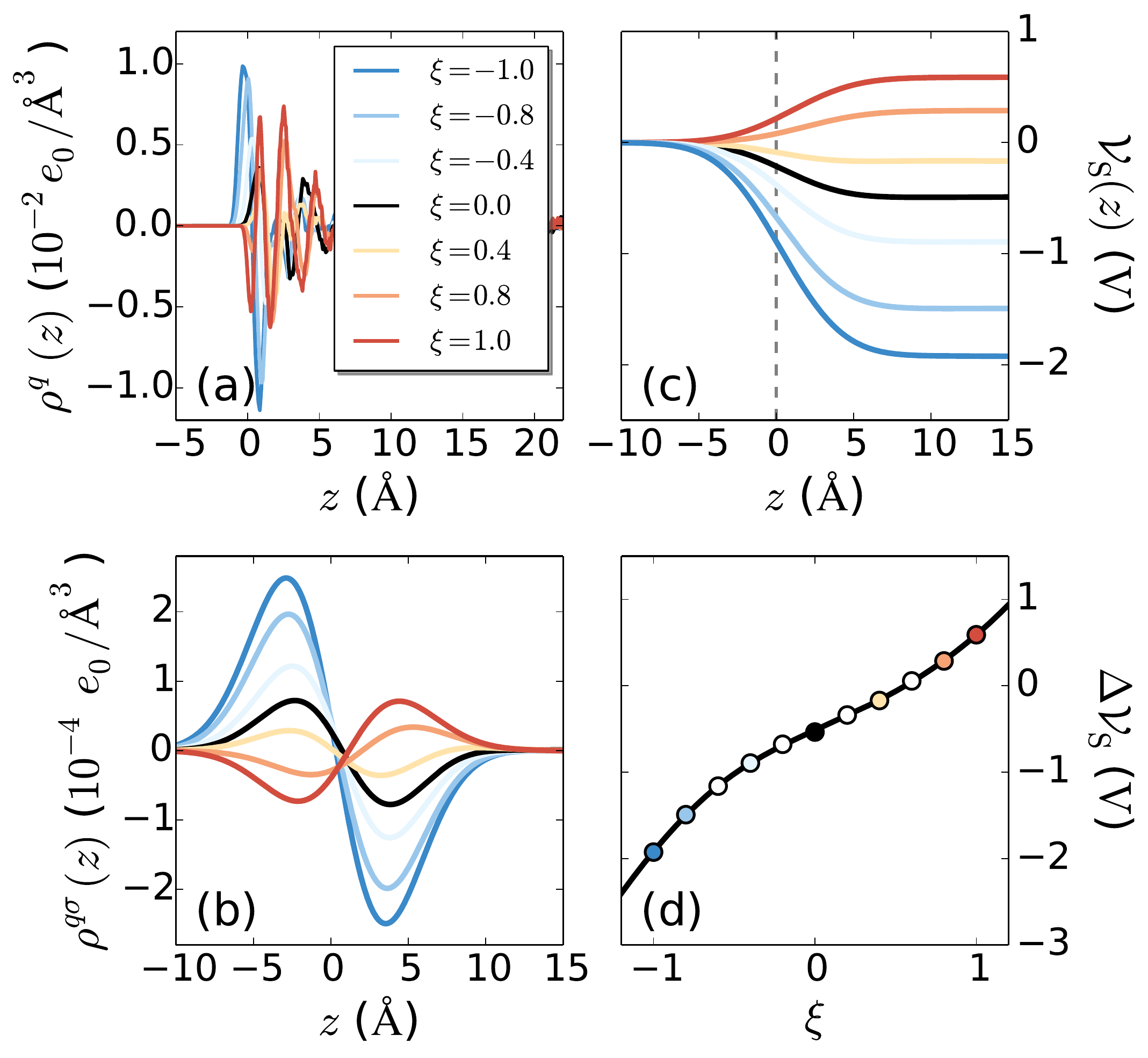}
\caption{(a) Bare charge densities provide little information about the collective electrostatic response of water to model silica surfaces of varying polarity, $\xi$.
(b) Upon Gaussian-smoothing, however, a long-ranged, collective dipole layer at the interface is apparent. This dipole layer characterizes the response of water to the polarity of the surface.
(c) The Gaussian-smoothed electrostatic potential due to water provides a slowly varying metric quantifying this interfacial dipole layer.
(d) The difference $\Delta\V_{\rm S}$ in the long-ranged electrostatic potential in bulk water and in the effective vacuum inside the silica surface,  or at the surface itself as indicated by the dashed line, characterizes the nonlinear response of water to the magnitude of the dipole of the surface.
}
\label{fig:1d}
\end{figure}

We first introduce our approach and its underlying concepts
through the study of uniform planar surfaces with varying hydrophobicity.
In particular, we study the response of water to the model atomically-detailed and corrugated silica-like surfaces of
Giovambattista, Rossky, and Debenedetti~\cite{Giovambattista:PNAS:2009,Giovambattista:PNAScorr:2013}.
These artificial surfaces are dipolar, with the dipole moment of surface groups created by negative partial charges placed on the uppermost
layer of oxygen sites and positive charges on the subsequent layer of silicon atoms.
In order to tune the relative polarity of the surface, the magnitude of the atomic charges, $q$, is linearly coupled
to a parameter $\xi$, such that the dipole moment of each surface unit is given by
$\xi\mathbf{p}=(0,0,\xi q d_{\rm OSi})$, where $d_{\rm OSi}$ is the oxygen-silicon bond length.
A nonpolar, hydrophobic surface with a contact angle of $\theta_\xi \approx109\degree$ is obtained when $\xi=0$.
Increasing the magnitude of $\xi$ leads to an increase in the hydrophilicity of the surface, as characterized by a decrease in the contact angle $\theta_\xi$, and there is a slight asymmetry with respect to the sign of $\xi$.
For example, $\theta_1\approx90\degree$, while $\theta_{-1}\approx85\degree$~\cite{Giovambattista:PNAS:2009,Giovambattista:PNAScorr:2013}.

In Figure~\ref{fig:1d}a, we show $\rhoq(z)$, the variation in the $z$-direction of the bare charge density of water,
\begin{equation}
\rho^q(\rb)\equiv \avg{\sum_i^{N_{\rm C}} q_i \delta(\rb-\rb_i(\Rbar))},
\end{equation}
where the sum is over all $N_{\rm C}$ charged sites with charge $q_i$
and position $\rb_i(\Rbar)$ in configuration $\Rbar$, and $\avg{\cdots}$
indicates an ensemble average over all configurations and
we have averaged over the remaining two coordinates in $\rb$.
Because of the rapidly varying molecular scale surface structure and charge distributions,
little insight into the collective response of water to the surface is garnered from $\rhoq(z)$,
and an alternative description of the ordering of interfacial water is needed.

\begin{figure*}
\centering
\includegraphics[width=0.8\textwidth]{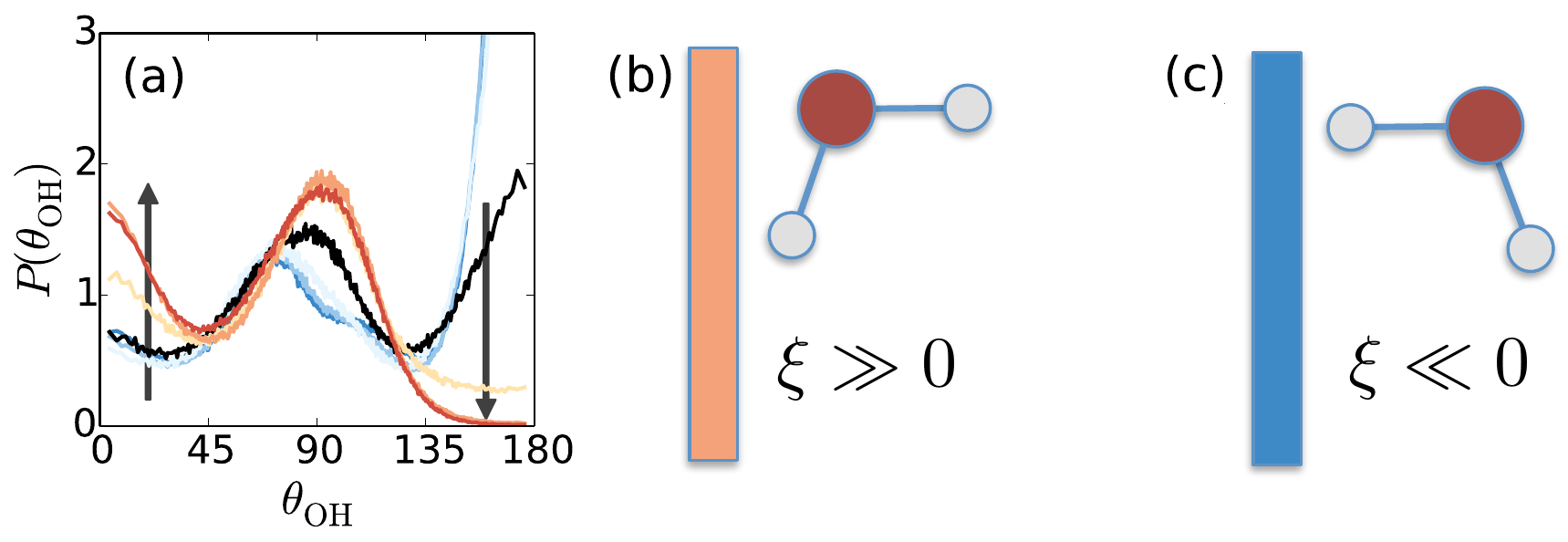}
\caption{
(a) The probability distribution of the angle made by the surface normal and the OH bond vector, $P(\theta_{\rm OH})$,
illustrates that water polarizes differently in response to surfaces with positive and negative $\xi$.
Gray arrows indicate the direction of increasing $\xi$.
(b) For $\xi\gg0$, interfacial water molecules tend to point an OH bond toward the bulk, while
(c) interfacial waters point an OH group toward the surface for $\xi\ll0$.
A small portion of each surface of varying $\xi$ is indicated by the colored rectangle, while the oxygen site is colored red and the hydrogens are colored light gray.
}
\label{fig:POH}
\end{figure*}

A clearer measure of the long-wavelength electrostatic response of water
induced by a heterogeneity is given by the Gaussian-smoothed
charge density $\rhoqs(\rb)$ that naturally arises within
the context of the local molecular field (LMF) theory of electrostatics~\cite{LMFDeriv}.
Here
\begin{equation}
\label{eq:Gausssmooth}
\rhoqs(\rb)=\int d\rb' \rho^q(\rb')\rhoG(\len{\rb-\rb'}),
\end{equation}
$\rhoG(\rb)$ is a normalized Gaussian function given by
\begin{equation}
\rhoG(\rb)=\frac{1}{\pi^{3/2}\sigma^3}e^{-\para{\frac{\len{\rb}^2}{\sigma^2}}},
\end{equation}
and $\sigma$ is a smoothing length used to average over short-ranged rapidly varying components of the electrostatic interactions.
Previous work has shown that
useful smoothing results when $\sigma$ is chosen on the order of nearest-neighbor distances~\cite{LMFDeriv,LMFWater,MolPhysLMF},
and we choose $\sigma=4.5$~\AA \ herein.
This type of smoothing over molecular sizes,
often discussed in elementary electrostatic texts~\cite{Jackson,Zangwill}, 
washes out the effects of atomic-scale details,
and it has been argued that $\rhoqs(\rb)$ provides a good qualitative characterization
of the underlying long wavelength electrostatic response of fluids~\cite{LMFWater,MolPhysLMF}.
Note that in the present case when the bare charge density $\rho^q(\rb)$ is known directly from simulation,
LMF theory is used only to motivate the consideration and usefulness of the smoothed charge density
$\rhoqs(\rb)$.


Indeed, as  shown in Figure~\ref{fig:1d}b, the Gaussian-smoothed charge densities readily
indicate the formation of an interfacial dipole layer arising from subtle changes
in the orientation of water molecules near the surface as they are increasingly pinned by the partial charges of the model surface.
The formation of a dipole layer is apparent even at the nonpolar surface with $\xi=0$, as detailed in previous work~\cite{RosskyJCP,LMFWater,LeeMR}. 

As $\xi$ is increased from zero in the negative direction, this interfacial dipole grows in magnitude.
The dominant molecular orientation producing the interfacial dipole for $\xi\ll0$ corresponds to that sketched in Figure~\ref{fig:POH}c. 
Here a non-negligible fraction of interfacial water molecules point one O-H bond directly toward the surface, and form hydrogen bonds with the negatively charged ``silicon'' atoms.
The collective interfacial dipole at $\xi=0$ arises from
the smaller fraction of water already existing in this orientation,
although most water molecules at the interface then have an orientation where
their H-O-H plane is parallel to the surface itself, which makes no contribution to the dipole layer~\cite{Trudeau:2009}.

Conversely, as $\xi$ is increased from zero in the positive direction, the magnitude of the interfacial dipole
decreases and eventually changes sign. This is consistent with interfacial water molecules
pointing an O-H bond away from the interface with increasing probability
as $\xi$ is increased, as evidenced by the 
probability distributions $P(\theta_{\rm OH})$ of
the angle made by the OH bond vector of a water molecule and the surface normal
shown in Figure~\ref{fig:POH}a.

Although the Gaussian-smoothed charge density has provided a physically suggestive description of the response of water to a uniform planar surface,
a more quantitative metric is needed to characterize patterned and corrugated surfaces and protein complexes.
To that end, we focus attention on $\V_{\rm S}(\rb)$,
the slowly-varying, long-wavelength portion of the electrostatic
potential due to the solvent. This is related to the smoothed charge density
$\rhoqs(\rb)$ by Poisson's equation
\begin{equation}
\label{eq:poisson}
\laplac \V_{\rm S}(\rb) = -\rhoqs(\rb),
\end{equation}
with notation chosen to be consistent with previous work~\cite{LMFDeriv}.
This potential, shown in Figure~\ref{fig:1d}d, smoothly transitions from zero in 
the vacuum region well inside the silica surface to
a value of $\Delta\V_{\rm S}$ in the bulk region.
Note that trends in $\Delta\V_{\rm S}$ can be evaluated using other reasonable limits, like the difference between the bulk potential
and the dashed line in Figure~\ref{fig:1d}c at the silica surface, without qualitatively affecting the characterization of the interface.
We use this fact below when generating an electrostatics-based map of a water-protein interface.

\begin{figure*}
\centering
\includegraphics[width=0.95\textwidth]{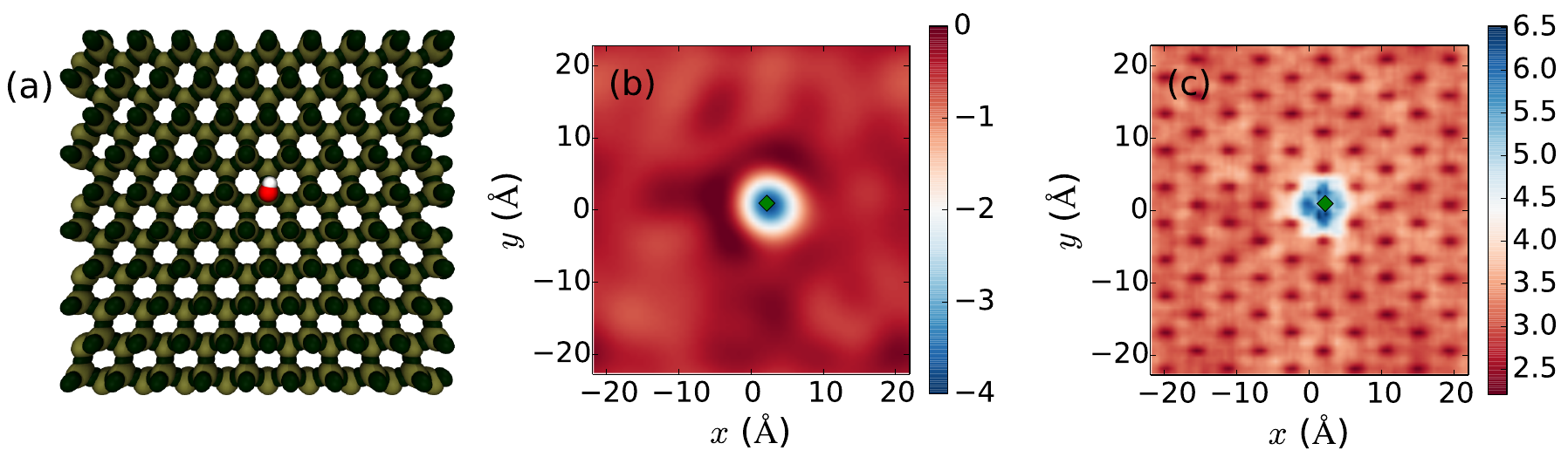}
\caption{
(a) An otherwise apolar silica surface, composed of oxygen (green) and silicon (yellow) atoms, is patterned with a single, hydrophilic hydroxyl group,
shown in red (oxygen) and white (hydrogen).
(b) The potential difference $\Delta\V_{\rm S}(x,y)$ (in Volts) from the three-dimensionally Gaussian smoothed charge density reveals that the polar and nonpolar portions of the surface
polarize nearby water differently, enabling the mapping of surfaces with complex chemistries.
(c) The characterization of this patterned surface provided by the solvation free energy of a hard cuboidal solute, $\beta\Delta\mu(x,y)$ (see text for details), illustrates that the
 electrostatics-based map is consistent with previous, water density fluctuation-based surface mapping techniques.
 A green diamond indicates the position of the hydroxyl group oxygen in (b) and (c).
 }
\label{fig:1hphil}
\end{figure*}

\begin{figure} [b]
\centering
\includegraphics[width=0.49\textwidth]{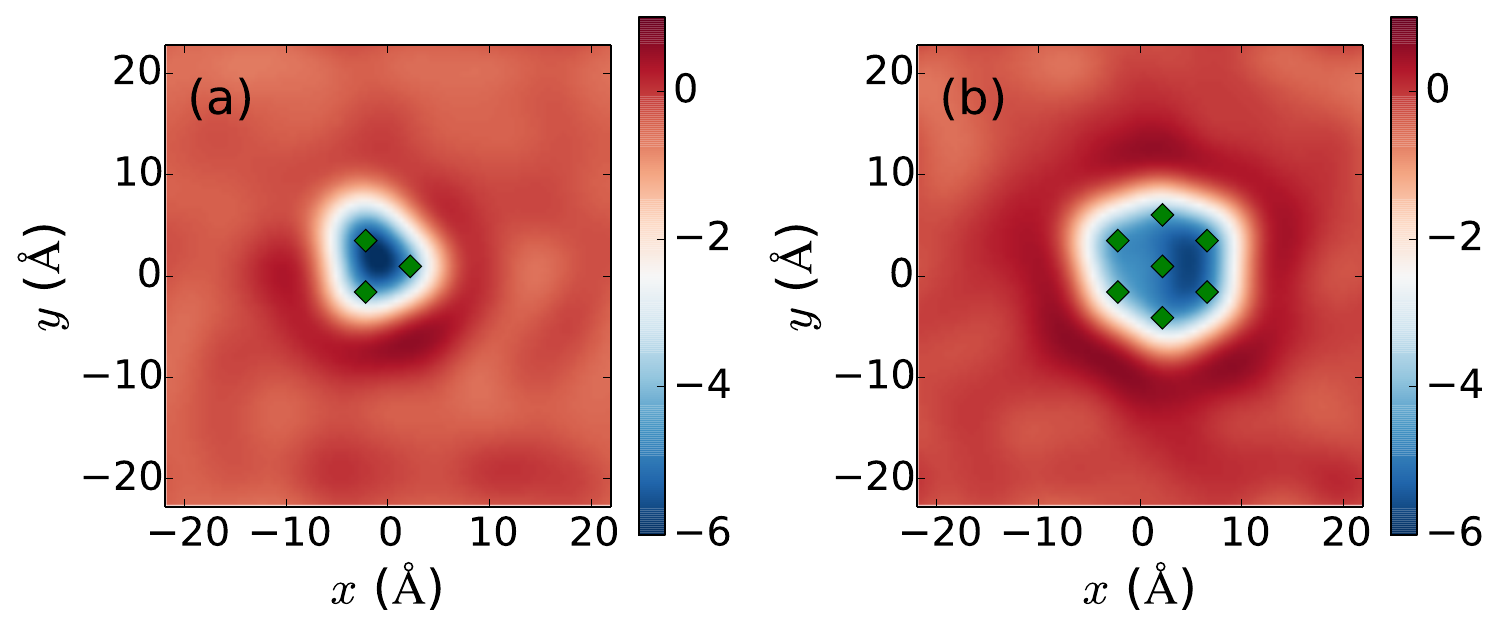}
\caption{
Water molecules at the surface are significantly polarized when (a) three and (b) seven hydrophilic (hydroxyl) groups
are added to an otherwise hydrophobic model silica surface, as indicated by the $\Delta \V_{\rm S}(x,y)$ plots shown here (in Volts).
The positions of the hydroxyl group oxygen sites are indicated by the green diamonds.
 }
\label{fig:bighphil}
\end{figure}

We propose that $\Delta\V_{\rm S}$ can be utilized as a quantitative
measure of the response of water to complex surfaces.
Indeed, $\Delta\V_{\rm S}(\xi)$ shown in Figure~\ref{fig:1d}d
quantifies the nonlinear response of the mesoscopic surface dipole layer to changes in $\xi$, such that the magnitude
and sign of water polarization at the surface is captured.
Additionally, the relative differences in $\Delta \V_{\rm S}(\xi)$ are due exclusively to structural rearrangements near the surface
(dipolar contributions), because the constant quadrupolar (Bethe potential) contribution to the potential is the same for all $\xi$-values~\cite{Remsing:2014}.

Moreover, recent work has emphasized that polar hydrophilic surfaces, which locally pin water molecules, are fundamentally different
from nonpolar hydrophilic surfaces that have artificially large Lennard-Jones-like attractions.
These latter surfaces simply pull interfacial waters closer to the surface, increasing the contact angle, but do not introduce significant changes in water
structure about the position of this interface~\cite{Willard:2014}.
Therefore, the local pinning of interfacial waters through polar interactions like hydrogen bonding underlies the hydrophilicity of realistic surfaces,
and the electrostatic mapping introduced here captures this behavior.

Finally, we note that in the simple case of a system with a slab-like geometry,
the potential difference obtained from the one-dimensional $\V_{\rm S}(z)$, $\Delta \V_{\rm S}$,
is exactly equal to that estimated from the bare electrostatic potential $\V(z)$, $\Delta \V$.
This equality is due to the fact that Gaussian smoothing of the potential conserves the first two nonzero multipole moments of a charge distribution (see the Appendix),
the dipole and quadrupole in the case of water, and the potential difference across two phases in a slab geometry depends only on these two moments
for nonionic systems~\cite{PrattComment,PrattReview}.
However, this is true only in the special case of slab-like symmetry.
Gaussian-smoothing over molecular length scales is critical for
quantifying the electrostatic response to complex surfaces, where $\V_{\rm S}(\rb)$ is three-dimensional and has no apparent symmetry, as we discuss below.



\section{Characterization of chemically patterned surfaces}

The above analysis can be readily generalized to more complex, patterned surfaces with less symmetry.
Here we show that the potential difference as a function of lateral position, $\Delta\V_{\rm S}(x,y)$, can be used to characterize the long-wavelength perturbations of the H-bond
network induced by local changes in the chemical patterning of a planar molecular surface.
Gaussian smoothing of the three-dimensional charge density in Eq. (\ref{eq:Gausssmooth}) ensures
that relevant non-local perturbations of the water structure due to patterning of the surface in the $xy$-plane are captured,
in addition to those in the $z$-direction, parallel to the surface normal

\begin{figure*}[t]
\centering
\includegraphics[width=0.95\textwidth]{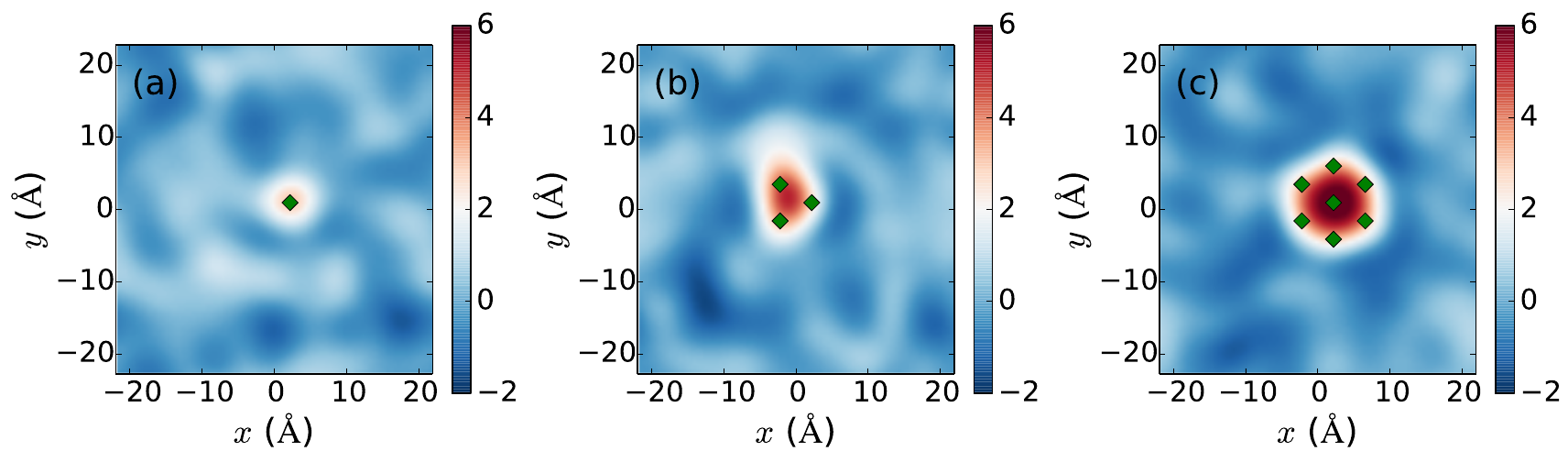}
\caption{
The electrostatics-based mapping provided by $\Delta\V_{\rm S}(x,y)$ (V) additionally provides a characterization of the response of water
to hydrophilic silica surfaces with (a) one, (b) three, and (c) seven hydrophobic, apolar units, the positions of which are indicated by the diamonds.
 }
\label{fig:hphobpattern}
\end{figure*}

We first consider a purely hydrophobic (uncharged) silica surface, and add a single
hydrophilic site, consisting of a hydrogen-bonding hydroxyl group~\cite{Giovambattista:PRE:2006,Giovambattista:JPCC:2007}, as shown in Figure~\ref{fig:1hphil}a.
A clear picture of the electrostatic response of water emerges upon examination of $\Delta\V_{\rm S}(x,y)$.
Long-wavelength perturbations of the water H-bond network extending over roughly 10~\AA \ in the plane of
the surface are found when a single hydrophilic site is added to the otherwise hydrophobic surface, Figure~\ref{fig:1hphil}b.
This is consistent with the concept that a single hydrophilic site can substantially pin water in its vicinity~\cite{Giovambattista:JPCC:2007,Acharya:Faraday:2010},
and also is in accord with the results detailed above for homogeneous surfaces.
As the number of hydrophilic sites is increased to three and seven sites, an increasingly larger region of 
water is perturbed, as illustrated in Figures~\ref{fig:bighphil}a and~\ref{fig:bighphil}b, respectively.

\begin{figure*}[t]
\centering
\includegraphics[width=0.95\textwidth]{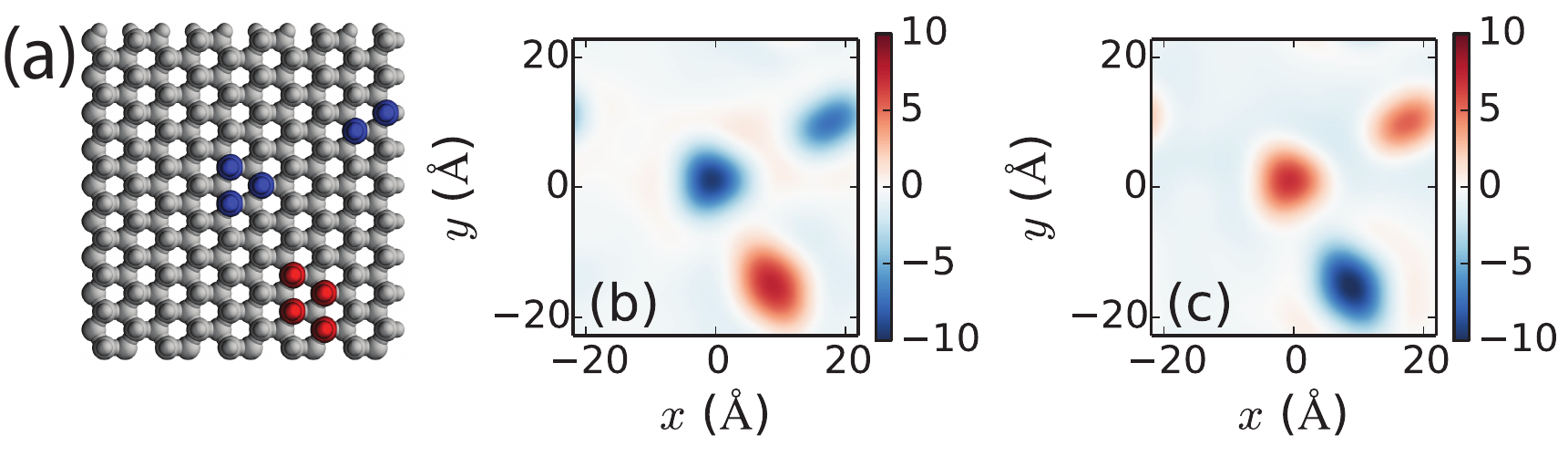}
\caption{(a) Snapshot of the model surface where the dipole moment of the surface
units in the polar regions colored red and blue
have opposite sign, while the rest of the surface is apolar (gray).
(b) Potential difference $\Delta\V_{\rm S}(x,y)$ due to the
Gaussian-smoothed charge density above the patterned silica surface.
(c) $\Delta \V_{\rm S}(x,y)$ obtained above the inverse of the surface in panel (a), where the signs of the dipoles
on the polar regions have been switched. Note that the opposite surface dipole moments induce differing responses in nearby water molecules,
as captured by the smoothed electrostatic potential difference shown here.
}
\label{fig:randpattern}
\end{figure*}

We also consider surfaces complementary to those in Figures~\ref{fig:1hphil} and~\ref{fig:bighphil},
created by replacing hydrophilic groups with hydrophobic groups, and vice versa.
For example, the complement of the surface in Figure~\ref{fig:1hphil}a has a single apolar unit at the center of a surface covered in hydroxyl groups. 
The electrostatics-based maps for these hydrophilic surfaces with one, three, and seven
hydrophobic sites are shown in Figure~\ref{fig:hphobpattern}a,~\ref{fig:hphobpattern}b, and~\ref{fig:hphobpattern}c, respectively.
Our findings are consistent with the idea that a hydrophobic site does not perturb the H-bond network near a hydrophilic surface nearly as much as a hydrophilic site
does on an otherwise hydrophobic surface~\cite{Acharya:Faraday:2010}.
This can clearly be observed by comparing Figure~\ref{fig:1hphil}b to Figure~\ref{fig:hphobpattern}a, where a single hydrophobic site
hardly perturbs the structure of water at the surface.

We additionally note that the response of water to a small hydrophobic patch on a hydrophilic surface is not equal to that above a uniform hydrophobic surface
(a similar statement holds for the complementary surfaces).
This equality becomes true for large enough patches, but the collective effects due to maintaining the H-bond network across the patch boundaries restricts the orientation
of interfacial water molecules, and leads to larger values of $\Delta \V_{\rm S}(x,y)$ above these small patches than is expected from uniform surfaces with the same chemistry, see Figure~\ref{fig:1d}d.

We further compare our electrostatics-based mapping procedure with a
density fluctuation-based mapping in Figure~\ref{fig:1hphil}c.
The fluctuation maps shown here were generated by calculating the probability $P_v(N;x,y)$ of observing $N$ water molecules
in a probe volume $v=\sigma\times \sigma\times 4$~\AA \ centered at the first peak of the nonuniform density
of water at the surface for all $x$ and $y$.
The free energy of emptying this volume over the surface is then given by $\beta\Delta\mu(x,y)=-\ln P_v(0;x,y)$
and this is used to probe the relative hydrophobicity of the surface.

This fluctuation-based mapping provides a description of interfacial water in accord with the electrostatics-based approach presented herein.
For all surfaces studied here, consistency is found between the fluctuation-based and electrostatic-based maps of relative hydrophobicity.

Although both fluctuation-based and electrostatics-based maps discriminate well between hydrophobic and hydrophilic regions of a surface,
the electrostatics-based mapping provides additional useful information about 
the polarization of water at the hydrophilic portions of the surface that standard fluctuations-based methods do not resolve.
To illustrate this point, we construct a model surface with regions of differing polarity shown in Figure~\ref{fig:randpattern}a.
The first patch (red) has four surface units with $\xi=1$, while the other two (blue) correspond to units with $\xi=-1$, one with three units and one with two.
We expect that water in the vicinity of the surface will be polarized in opposite directions above the different dipolar units.
The remainder of the surface is nonpolar (gray).

We find that $\Delta \V_{\rm S}(x,y)$ can readily distinguish between the regions of differing polarity.
Water in the vicinity of the $\xi=1$ patch responds in a manner that yields a positive $\Delta \V_{\rm S}$,
while a negative $\Delta \V_{\rm S}$ is obtained over $\xi=-1$ regions. 
Surprisingly, the signs of these potentials are opposite to those obtained at uniform surfaces, stemming from the small size of the patches and the fact that most of the surface is apolar.
Water molecules above the apolar surface do not penetrate into the grooves of the surface, as do waters near uniform $\xi=1$ and $\xi=-1$ surfaces.
Penetration of water into the surface grooves near the small patches in Figure~\ref{fig:randpattern}a would lead to significant distortions of the hydrogen
bond network, which are not compensated by these weakly hydrophilic patches.
Instead, the orientation of water molecules in the vicinity of these patches is dominated by the tendency for their dipoles to align with the dipolar field of these patches.

This effect is due to the weakly hydrophilic nature of these artificial patches and the collective nature of the water hydrogen bond network.
We expect a transition in $\Delta \V_{\rm S}$ with patch size, until the infinite limit shown in Figure~\ref{fig:1d}d is reached.
Indeed, differences in $\Delta \V_{\rm S}$ with patch size are already observed in moving from two to three dipolar units (top right to central blue patches in Figure~\ref{fig:randpattern}a);
the magnitude of $\Delta \V_{\rm S}$ increases.
Such context-dependent behavior cannot be captured by focusing on the properties of the surface alone.
In this case, surface-based mapping would predict no difference between these patches and the uniform surface because they have the same basic charges, despite the very different response
of water based on the environment.
These results provide a dramatic illustration of both the flexibility and subtlety of the nonlocal response of the hydrogen bond network to various perturbations, and the ability of the electrostatic maps to resolve their effects.

We also consider the ``inverse'' of the surface in Figure~\ref{fig:randpattern}a, which has the same pattern, but the signs of the dipoles are opposite to those of the surface in Figure~\ref{fig:randpattern}a.
Thus, we expect water in the vicinity of these patches to be polarized in the direction opposite to those of the inverse surface.
Indeed, the electrostatic mapping described by $\Delta \V_{\rm S}(x,y)$ demonstrates this to be true, as shown in~\ref{fig:randpattern}c.
This illustrates that the electrostatic mapping of interfaces readily uncovers the qualitatively different polarization of water at polar surface units,
in addition to the relative hydrophobicity of regions of the surface.


\section{Mapping protein surfaces}

The surface of the protein Hydrophobin II (HFBII) has both a hydrophilic and a hydrophobic region, allowing the protein to effectively act as a biomolecular Janus particle
and assemble at hydrophobic-aqueous interfaces~\cite{Hakanpaa:2006ab}.
HFBII has been analyzed using classic hydrophobicity scales, in addition to the water density fluctuation-based techniques
described in the Introduction~\cite{Acharya:Faraday:2010,Patel:JPCB:2014}. 
The wide array of chemical and topographical complexities on the HFBII surface thus presents an ideal test case
for extending the electrostatic mapping technique to protein surfaces.

In contrast to simple planar interfaces, one cannot generally define a vacuum region on the protein side of the protein-water interface.
Instead, we can define a water-protein interface, $\sbb$, and evaluate the difference between the value of the long-ranged potential in the bulk and at this interface, $\Delta \V_{\rm S}(\sbb)$.
This modified metric can still distinguish between regions of varying polarity, since our qualitative findings for planar interfaces would be unaltered by shifting the vacuum reference
to the position of the interface ($z\approx0$~\AA \ in Figure~\ref{fig:1d} for example).

In the results shown here, we follow previous work~\cite{Patel:JPCB:2014} and evaluate $\Delta \V_{\rm S}(\sbb)$ at the Willard-Chandler smoothed interface defined by the protein heavy atoms alone~\cite{InstInt}.
We choose a low value of the density to define the interface, so that this surface envelopes the heavy atoms of the protein and is similar to a typical solvent excluded surface used commonly throughout the literature.  
As illustrated in Reference~\citen{LMFDeriv}, because of the Gaussian smoothing in $\V_{\rm S}(\rb)$, many molecular details in the protein substrate are unimportant.
Thus the surface mapping procedure is rather insensitive to the precise definition of the protein-water interface, and many other reasonable choices could be made as well.

\begin{figure}[t]
\centering
\includegraphics[width=0.49\textwidth]{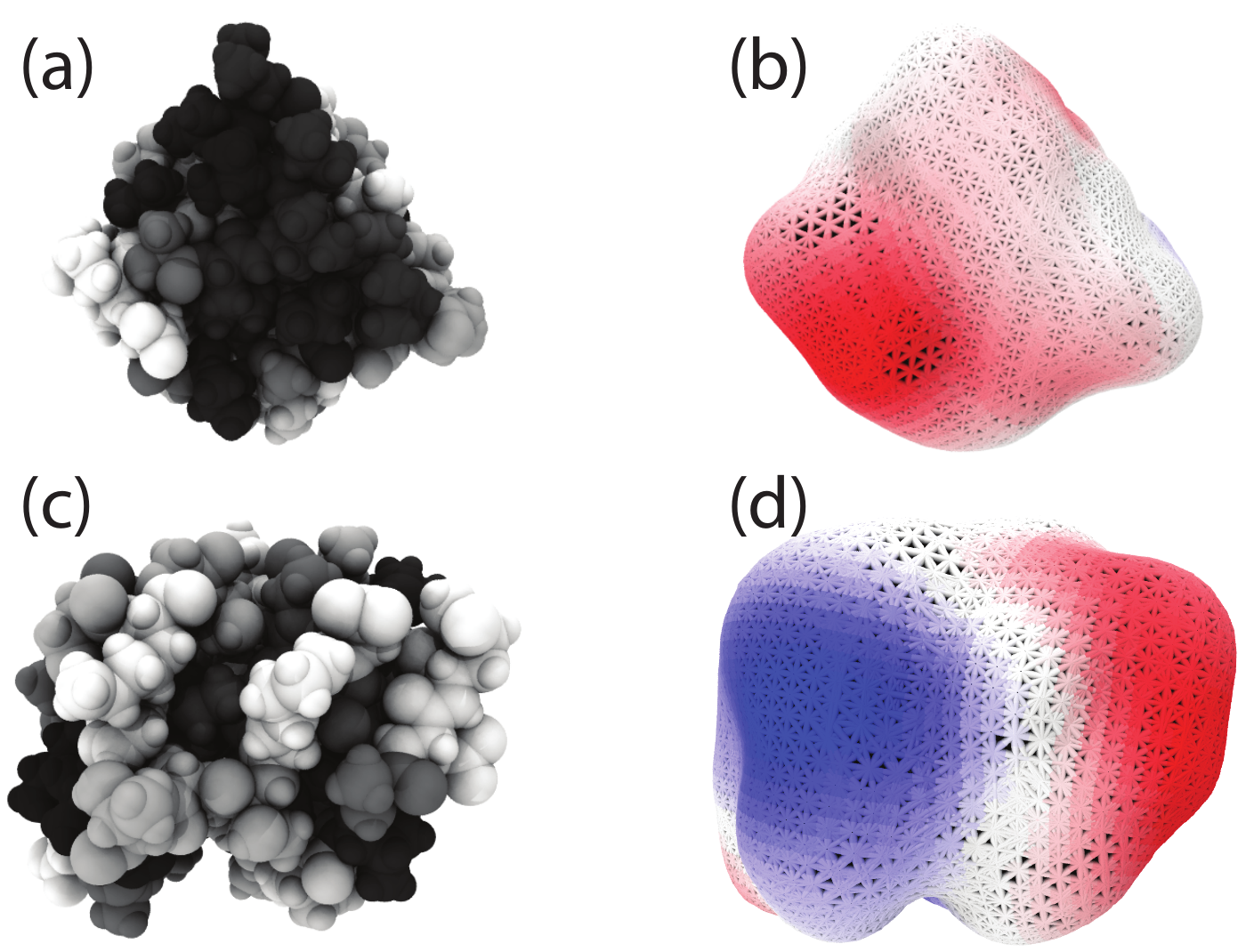}
\caption{
(a,c) Two views of the Kyte-Doolittle residue-based mapping of Hydrophobin II (HFBII); increasing hydrophobicity coincides with the darkness of the residue.
Panel (a) highlights large hydrophobic patch on the surface of the protein, while panel (c) focuses on hydrophilic regions.
(b,d) The analogous two views of the electrostatic mapping of the water-HFBII interface.
The protein surface as described by a Willard-Chandler smoothed interface for the protein heavy atoms is shown as a meshgrid and is colored by value of the $\Delta \V_{\rm S}(\sbb)$.
Red represents negative values of the potential, blue indicates positive values of the potential, and white corresponds to values near zero, such that slightly negative, pale red/pink indicates values
of the $\Delta \V_{\rm S}$ observed above hydrophobic surfaces. Figures were made using VMD~\cite{HUMP96}.
}
\label{fig:prot}
\end{figure}

The electrostatics-based mapping of HFBII is compared to the classic Kyte-Doolittle hydrophobicity scale in Figure~\ref{fig:prot}.
The Kyte-Doolittle scale is indicated by the color of the residues in Figure~\ref{fig:prot}a and~\ref{fig:prot}b, such
that hydrophobicity is proportional to the darkness of the residue color.
The value of $\Delta \V_{\rm S}(\sbb)$ in Figure~\ref{fig:prot}b and~\ref{fig:prot}d is indicated by the color of the grid points on the mesh,
which corresponds to the Willard-Chandler surface of the protein.
The average interface is colored such that red represents large negative values of $\Delta \V_{\rm S}(\sbb)$,
blue indicates large positive values of the potential, and white corresponds to values near zero.
With this scaling, values of $\Delta \V_{\rm S}(\sbb)$ indicative of hydrophobic surfaces are colored pink/pale red.

The electrostatic mapping procedure readily uncovers the large hydrophobic patch of HFBII shown in Figure~\ref{fig:prot}a and~\ref{fig:prot}b.
The value of $\Delta \V_{\rm S}(\sbb)$ in the region of this patch is consistent with that for a hydrophobic surface.
However, note that there is a highly hydrophilic group at the edge of this hydrophobic patch (bottom left, red region). 
The polarization caused by this patch does not dissipate instantly, and the size of the hydrophobic patch is predicted to be smaller than that from surface-based scales.
This agrees rather well with findings of other water-based characterizations of HBFII~\cite{Acharya:Faraday:2010,Patel:JPCB:2014}.


As was detailed in the previous section, polar surfaces can polarize water in opposing directions.
The maps of HFBII indicate the formation of such patterns in a  biological context.
The blue (positive $\Delta \V_{\rm S}(\sbb)$) portion of the map in Figure~\ref{fig:prot}b is dominated by the response of water molecules to the positively charged
lysine residue in the region.
Analogously, the red (negative $\Delta \V_{\rm S}(\sbb)$) region on the right of Figure~\ref{fig:prot}b describes the response of water to a negatively charged aspartic acid residue.
The identification of such hydrophilic regions of differing polarity may be crucial to understanding the binding of biomolecules.
For example, these oppositely charged residues come into close contact with those of neighboring HFBII molecules when forming multimeric assemblies~\cite{Hakanpaa:2006aa},
and a characterization of the polarization of water near polar regions of proteins may help predict their involvement in biomolecular interactions~\cite{Sheinerman:2000ab}.

\section{Conclusions}

In this work, we have introduced a novel electrostatics-based method for characterizing the long-wavelength, collective response of water to complex surfaces.
This approach effectively coarse-grains over molecular scale details in order to uncover the underlying electrostatic behavior of the system, enabling
one to distinguish larger scale rearrangements of the hydrogen bond network of water at an interface and consequently differentiate hydrophobic and hydrophilic surfaces.
This approach is quite efficient computationally, requiring only a single equilibrium simulation of a complex surface in explicit water.


In addition to distinguishing between hydrophobic and hydrophilic regions, the approach introduced here can further distinguish the relative polarization of water in response to hydrophilic groups.
An understanding of the relative polarization of water at surfaces is important for a range of processes, including understanding and predicting ion adsorption to biomolecular interfaces.
It has been postulated that the sign of the surface potential can drive ions to adsorb to interfaces, at least in classical models~\cite{BaerDCT}.
As such, if adsorption were to be favorable at a patterned surface like that in Figure~\ref{fig:randpattern}, for example,
one might expect that ions of opposite sign would adsorb to regions of opposite dipole moments.

Differentiating the relative polarity of water in interfacial regions may also be useful in the description of binding processes in general.
In particular, it seems plausible that polarization-based hydration repulsion could occur between two surfaces~\cite{Schneck:2012aa,Kanduc:2014}.
For example, consider placing in close proximity two surfaces that have maps like those in Figure~\ref{fig:randpattern}b, such that the blue regions align.
As the two surfaces are brought closer together, polarized interfacial water may ``interfere deconstructively'' and effectively repel the two surfaces, because the hydrogen bond network
is polarized in different directions everywhere but the midpoint between the interfaces, where the polarization must vanish by symmetry~\cite{Schneck:2012aa,Kanduc:2014}.
Therefore, to achieve small interplate distances, confined water must be (unfavorably) depolarized.

In contrast, one might expect that bringing together surfaces with regions of opposing polarity, like the red and blue regions in Figure~\ref{fig:randpattern},
would instead lead to hydrogen bond networks that interfere \emph{constructively}, because the inter-surface water is polarized in the same direction everywhere between these patches.
In this case, unfavorable depolarization may not occur, minimizing this contribution to the effective repulsion between surfaces and possibly leading to an effective attraction
between hydrophilic surfaces~\cite{Kanduc:CPL:2014,Kanduc:2014}.

However, in more general cases of molecular assembly, the polarization of water molecules, and therefore $\Delta \V_{\rm S}(\rb)$,
depends on a multitude of parameters including the separation and relative orientation of the confining surfaces,
and more than bulk equilibrium simulations  of individual surfaces are required for a quantitatively accurate description of such phenomena.
Nevertheless, the electrostatic mapping described herein should provide qualitative insight into many relevant features of solvent polarization effects needed to understand hydration mediated forces.



\section{Methods}
\subsection{Simulation of Silica Surfaces}

All simulations of water near model silica surfaces were performed using a modified version of the DL\_POLY software package (version 2.18)~\cite{dlpoly},
following the work of Hu and Weeks~\cite{Hu:JPCC:2010}.
Simulation cells $45.6\times43.9\times180$~\AA$^{3}$ in volume were used to simulate 2468 SPC/E~\cite{SPCE} water molecules at model silica surfaces with a buffering liquid-vapor
interface far from the surface, to maintain a constant coexistence pressure while in the canonical ensemble
(The large size of the cell in the $z$-dimension is also useful for evaluating electrostatics in slab like geometries.).
A temperature of 298~K was maintained using a Berendsen thermostat~\cite{BerendsenBaroThermo}.
Lennard-Jones interactions were truncated and shifted at a distance of 11~\AA.
The slab corrected Ewald summation method of Yeh and Berkowitz~\cite{Ewald3DC}
was used to handle the evaluation of electrostatic interactions, with a real space cutoff of 11~\AA, switching parameter $\alpha=0.3$~\AA$^{-1}$, and
a maximum number of $k$-space vectors of $k_x=10$, $k_y=10$, and $k_z=30$ in the $x-$, $y-$, and $z-$directions, respectively,
such that the $k$-space sums run from $-k_x$ to $k_x$, for example.
For the patterned dipolar surfaces discussed in Figure~\ref{fig:randpattern}, a larger number of $k$-vectors is needed, such that
$k_x=20$, $k_y=20$, and $k_z=30$ for these simulations.

\subsection{Protein Simulation}

Simulations of HFBII (PDB ID: 2B97)~\cite{Hakanpaa:2006ab} in SPC/E~\cite{SPCE} water were performed using the GROMACS 4.5.3 software package~\cite{gmx4ref}
and the AMBER 94 force field~\cite{AMBER}, following the work of Patel and Garde~\cite{Patel:JPCB:2014}.
As with the silica surfaces, simulations were performed in the presence of a buffering liquid-vapor interface, now with one in each $z$-direction.
The protein atoms were held fixed at the center of the liquid slab.
Production simulations used for analysis totaled 6~ns in length, and a constant temperature of 300~K was maintained using the canonical velocity rescaling algorithm~\cite{Bussi:JCP:2007}.
All Lennard-Jones interactions were truncated at a distance of 10~\AA, and long-ranged corrections to the energy and pressure were ignored.
Electrostatic interactions were evaluated using the particle mesh Ewald method~\cite{PME}.

\begin{acknowledgements}
This work was supported by the National Science Foundation (Grants CHE0848574 and CHE1300993).
We are grateful to Jocelyn Rodgers for helpful discussions.
RCR also acknowledges Amish Patel for stimulating discussions and Erte Xi for assistance with the generation of the protein interface. 
\end{acknowledgements}

\appendix
\section{Multipole Moment Expansion of Gaussian-Smoothed Charge Densities}
\label{sec:appendixA}
Here, we present the multipole expansion for the Gaussian-smoothed charge density,
$\rhoqs(\rb)$, and how it relates to that of the bare charge density $\rho^q(\rb)$.
For generality, we consider the interaction energy between two $d$-dimensional Gaussian-smoothed
charge densities, with centers $\rb_i$ and $\rb_j$, where the position vector is defined by
\begin{equation}
\rb=\para{x_1,x_2,...,x_d},
\end{equation}
and $d\in\mathbb{N}$.
We consider the interaction energy of the two charge distributions
under consideration, $\rhoqs_i(\rb_i)$ and $\rhoqs_j(\rb_j)$, respectively,
to be of the form
\begin{equation}
w(\rb_{ij})=\int d\rb \int d\rb' \rhoqs_i(\rb-\rb_i)\rhoqs_j(\rb'-\rb_j)\frac{1}{\epsilon \len{\rb-\rb'}},
\end{equation}
where we use the $3$-dimensional Coulomb potential.
However, the multipoles themselves depend only on the dimensionality of the charge density, and the relationship between the multipoles of the bare and smoothed charge densities
described below are independent of the form of the potential.
This interaction energy
can be rewritten as a $\kb$-space integral, 
\begin{equation}
w(\rb_{ij})=\frac{1}{(2\pi)^d}\int d\kb \frhoqs_i(-\kb) \frhoqs_j(\kb)e^{-i \kb \cdot \rb_{ij}}\frac{4\pi}{\epsilon k^2}, 
\label{eq:ftws}
\end{equation}
where we have defined the $d$-dimensional Fourier transform and inverse Fourier transform of a function $f$ as
\begin{equation}
\hat{f}(\kb)=\int d\rb e^{-i\kb \cdot \rb}f(r)\nonumber 
\end{equation}
and
\begin{equation}
f(\rb)=\frac{1}{\para{2\pi}^d}\int d\kb e^{i\kb\cdot\rb}\hat{f}(\kb),\nonumber
\end{equation}
respectively. 

Now to examine the asymptotic behavior as $k\rightarrow 0$, we Tayor expand
the smoothed charge densities about $ k=0$, such that
\begin{equation}
\frhoqs_i(\kb)=\sum_{n_i} \frac{1}{n_i !} \kb^{n_i} \cdot \gradk^{n_i} \frhoqs_i(0),\label{eq:exprhoqs}
\end{equation}
where
$\gradk$ is the $d$-dimensional gradient with respect to $\kb$.
We can then insert \ref{eq:exprhoqs} into \ref{eq:ftws} to obtain
\begin{align}
w(\rb_{ij}) &= \sum_{n_i,n_j}\frac{1}{n_i ! n_j !} \brac{\gradk^{n_i}\frhoqs_i(0)\cdot (-i\gradr)^{n_i}} \nonumber \\
&\cdot\brac{\gradk^{n_j}\frhoqs_j(0)\cdot (i\gradr)^{n_j}}\frac{1}{\epsilon r}
\label{eq:expw}
\end{align}
Now, we can define
\begin{equation}
\frac{i^{n_i}}{n_i!}\gradk^{n_i}\frhoqs_i(0)\equiv\Mbs_i(n_i),
\end{equation}
such that 
\begin{equation}
\Mbs_i(n_i)=\frac{1}{n_i!}\int d\rb \rhoqs_i(\rb) \rb^{n_i},
\end{equation}
and $\Mbs(n)$ is the $n$th multipole moment of the smoothed charge distribution.
Finally, we can rewrite \ref{eq:expw} as 
\begin{equation}
w(\rb_{ij})=\sum_{n_i,n_j} \brac{\Mbs_i(n_i) \cdot (-\gradr)^{n_i}} \cdot\brac{\Mbs_j(n_j)\cdot \gradr^{n_j}}\frac{1}{\epsilon r}
\end{equation}
so that the energy is now expressed in terms of the multipole
moments of the smoothed charge distributions. 

One may then inquire into how these multipole moments
relate to those of the bare charge densities, $\rho^q(\rb)$.
In order to relate the two sets of multipoles, we first consider the
Fourier transform of the smoothed charge density, which,
using the convolution theorem, can be written as
\begin{equation}
\frhoqs(\kb)=\frho^q(\kb)\frhoG(\kb),
\end{equation}
where 
\begin{equation}
\frhoG(\kb)=e^{-\frac{k^2\sigma^2}{4}}.
\end{equation}

In general, the $n$th order multipole moment $\Mbs(n)$ is given by 
\begin{equation}
\Mbs(n)=\frac{i^{n}}{n!}\sum_{m=0}^n \binom{n}{m} \frho^{q(n-m)}(0)\otimes\frhoG^{(m)}(0),
\label{eq:genM}
\end{equation}
where $\hat{f}^{(n)}(0)=\brac{\gradk^n\hat{f}(\kb)}\bigg|_{k=0}$ is a tensor of rank $n$,
$\otimes$ indicates a symmetric outer product, and
\begin{equation}
\binom{n}{m}=\frac{n !}{m!(n-m)!}\nonumber
\end{equation}
is the binomial coefficient. 
The gradients of the $\kb$-space Gaussian function are given by
\begin{equation}
\frhoG^{(n)}(\kb)=(-1)^n e^{-k^2\sigma^2/4}\hermite{n}{\frac{\kb\sigma}{2}},
\end{equation}
such that $\hermite{n}{a\mathbf{x}}$ is a rank $n$ tensor-analog of the Hermite functions with elements
\begin{equation}
H_{ij\cdots v}(a\mathbf{x};n)=(-1)^n e^{a^2\mathbf{x}^2} \frac{\partial^n}{\partial x_i \partial x_j \cdots \partial x_v} \para{e^{-a^2\mathbf{x}^2}},\label{eq:hermite}
\end{equation}
where $a$ is a constant and $\xb=(x_1,x_2,...,x_d)$ is a general $d$-dimensional vector.


All odd derivatives of $\frhoG(\kb)$ will vanish at $k=0$ due to symmetry,
therefore, we can rewrite \ref{eq:genM} as
\begin{equation}
\Mbs(n)=\frac{i^n}{n!}\sum_{\substack{ m=0 \\ m\in\mathbb{E}}}^{n}(-1)^m\binom{n}{m}\frho^{q(n-m)}(0)\otimes\mathbf{A}_m\para{\frac{\sigma}{2}},
\label{eq:msign}
\end{equation}
where $\mathbf{A}_m(a)\equiv\brac{\hermite{m}{a \kb}}\big|_{k=0}$
and $\mathbb{E}$ is the set of even whole numbers.

Equation~\ref{eq:msign} can be written in the equivalent form
\begin{equation}
\Mbs(n)=\Mb(n)+\sum_{\substack{ m=2 \\ m\in\mathbb{E}}}^{n}\frac{(-1)^m i^m}{m!}\Mb(n-m)\otimes\mathbf{A}_m\para{\frac{\sigma}{2}},
\label{eq:lrmult}
\end{equation}
making the relation between $\Mbs(n)$ and $\Mb(n)$ apparent.
Therefore, in order for $\Mbs(n)=\Mb(n)$ to hold, where $\Mb(n)$ is the
$n$th multipole moment of the bare charge distribution $\rho^q$,
all multipoles of the bare charge density $\rhoq$ of order less than $n-1$ and of even (odd) order, for $n$ even (odd),
must be identically zero,
\begin{equation}
\Mbs(n)=\Mb(n) \iff \Mb(s)=0 \ \forall \ s=n-l,
\end{equation}
where
\begin{equation}
l= \bigg\{\begin{array}{ll}2,4,6,...,n; \ \text{for $n$ even}\\2,4,6,...,n-1; \ \text{for $n$ odd}.
\end{array}\end{equation}

To illustrate this condition, we present the first few multipole moments of the Gaussian-smoothed charge density.
The monopole moment of $\rhoqs$ is trivially given by $\Mbs(0)=\Mb(0)$, and note that for neutral charge distributions the monopole moment is zero.
The dipole moment, $n=1$, is also trivially given by
\begin{equation}
\Mbs(1)=\Mb(1),\nonumber
\end{equation}
illustrating that the dipole moment of $\rhoq$ is conserved upon Gaussian-smoothing.
In addition, the quadrupole moment is given by
\begin{eqnarray}
\Mbs(2)&=&\Mb(2)-\frac{1}{2}\Mb(0)\otimes\mathbf{A}_2\para{\frac{\sigma}{2}}\nonumber \\
&=&\Mb(2)+\frac{\sigma^2}{4}\Mb(0)\mathbf{I}_3,
\end{eqnarray}
while the Gaussian smoothed octupole is similarly given by
\begin{eqnarray}
\Mbs(3)&=& \Mb(3) + \frac{\sigma^2}{4}\Mb(1)\otimes\mathbf{I}_3,
\end{eqnarray}
where $\mathbf{I}_n$ is the $n\times n$ identity matrix.
For neutral charge distributions, like non-ionic molecular charge distributions (water),
both the dipole \textit{and} quadrupole moments are conserved upon Gaussian-smoothing, but not higher order moments.
From the work of Wilson and Pratt~\cite{PrattComment}, it follows that only the potential difference
across two phases in a slab-like geometry is conserved upon Gaussian-smoothing of the charge density (or the potential), because this
depends only on the dipole and quadrupole moments of a molecular fluid.
However, contrary to a previous report~\cite{Vorobyov:2010}, $\V_{\rm S}(z)\neq\V(z)$, because higher order multipoles are modified by smoothing.

\bibliography{jpcbelcHphob-rev3}



\end{document}